\documentclass[aps,pra,superscriptaddress,tightenlines,twocolumn,floatfix,notitlepage]{revtex4-2}
\usepackage{graphicx}
\usepackage[english]{babel}
\usepackage{amsmath}
\usepackage{amssymb}
\usepackage{tensor}
\usepackage{xcolor}    % color
\usepackage[T1]{fontenc}
\usepackage{mathptmx}
\DeclareMathAlphabet{\mathcal}{OMS}{cmsy}{m}{n}
\DeclareSymbolFont{Letters}{OML}{cmm}{m}{it}
\DeclareMathSymbol{\psi}{\mathalpha}{Letters}{32}
\DeclareMathSymbol{\Psi}{\mathalpha}{Letters}{9}
\usepackage[colorlinks,linkcolor=blue,citecolor=blue,urlcolor=blue]{hyperref}
\usepackage{cleveref}
\crefname{equation}{Eq.}{Eq.}
\crefname{figure}{Fig.}{Fig.}
\crefname{table}{Tab.}{Tab.}
\crefname{section}{Sec.}{Sec.}
\usepackage{float}

\newcommand{\be}{\begin{eqnarray}}
\newcommand{\ee}{\end{eqnarray}}

\newcommand{\dc}{c^{\dagger}}

\def\ket#1{|#1\rangle}

\newcommand{\beq}{\begin{equation}}
\newcommand{\eeq}{\end{equation}}
\newcommand{\ben}{\begin{align}}
\newcommand{\een}{\end{align}}
\newcommand{\bea}{\begin{aligned}}
\newcommand{\eea}{\end{aligned}}
\newcommand{\bes}{\begin{subequations}}
\newcommand{\ees}{\end{subequations}}
\newcommand{\bew}{\begin{widetext}}
\newcommand{\eew}{\end{widetext}}

\begin{document}

\title{Ring-Frustrated Non-Hermitian $XY$ Model}
\author{Shihao Bi}
\email{bishihao@stu.scu.edu.cn}
\affiliation{College of Physics, Sichuan University, 610064, Chengdu, People's Republic of China\\
and Key Laboratory of High Energy Density Physics and Technology of Ministry of Education, Sichuan University, 610064,
Chengdu, People's Republic of China}

\author{Yan He}
\email{heyan_ctp@scu.edu.cn}
\affiliation{College of Physics, Sichuan University, 610064, Chengdu, People's Republic of China\\
and Key Laboratory of High Energy Density Physics and Technology of Ministry of Education, Sichuan University, 610064,
Chengdu, People's Republic of China}

\author{Peng Li}
\email{lipeng@scu.edu.cn}
\affiliation{College of Physics, Sichuan University, 610064, Chengdu, People's Republic of China\\
and Key Laboratory of High Energy Density Physics and Technology of Ministry of Education, Sichuan University, 610064,
Chengdu, People's Republic of China}

\begin{abstract}
We study a non-Hermitian version of XY closed chain with odd number of lattice sites. We consider both anti-ferromagnetic coupling and also a symmetric non-collinear spin coupling. It is found that the energy spectrum is real in certain region of the parameter space. In contrast to previous non-Hermitian models, the ground state is a state with one mode occupied inside this real energy spectrum region, instead of the artificially identified vacuum state. At the same time, there appears a gapless excitation, which is made by kink like spin configurations. It is also found that this kink phase has non-trivial topological invariant.
\end{abstract}

% \pacs{
% 75.10.Jm % Quantized spin models, including quantum spin frustration
% 03.65.-w % Quantum mechanics
% 64.70.Tg % Quantum phase transitions
% 	  }
\maketitle

\section{Introduction}

In recent years non-Hermitian physics \cite{bender1998,bender2007} has attracted wide research interests
and growing enthusiasm for its intimate connection with open systems \cite{Carmichael1993,Rotter2009,Malzard2015},
optic systems with gain and loss \cite{Makris2008,Klaiman2008}, electron-electron interaction
or disorder induced self-energy in the effective Hamiltonian \cite{Shen2018,Fu2019}. New phenomenon \cite{Yao1} have
been discovered and the traditional band theory has been enriched and extended \cite{Yao2,Shen2018topological}.
A lot of efforts have been devoted to investigate the topological aspects of non-Hermitian system both theoretically and experimentally.
The non-Hermitian version of quantum spin chain has also been studied in previous works \cite{Zhang2013tim,Zhang2013xy,Zhang2013xy2,Song2015,Song2016}.

In the quantum spin model with anti-ferromagnetic exchange interaction and odd number of lattice sites, there appears some interesting nonlocal effects due to the so-called ring frustration \cite{dong2016,dong2017}. The ring frustration is a type of geometrical frustration caused by the impossibility of accommodating the staggered spin configuration on a closed spin chain with odd number of sites. Because of this frustration, the low energy spin configuration always contains a kink like structure, which give rise to a gapless excitation in the anti-ferromagnetic phase. It is also found that ring frustrations also lead to nonlocal correlation functions in a variety of spin models \cite{dong2018,li2019,he2020}. Recently, similar ideas have been generalized to other exactly solvable interacting fermion model in \cite{zheng2019}.

In this paper, we study the XY model with anti-ferromagnetic coupling and also a non-Hermitian interaction, which is an exactly solvable non-Hermitian spin model. One may wonder if the ring frustration also apply to this non-Hermitian model when the lattice site number is odd. Because the interaction energies here are complex number, one may feel that the staggered spin configuration may not be favored and the effects of ring frustration will not occur. Actually, we will show that in certain region of the parameter space, the eigenvalues of the Hamiltonian are real, and in this area the ring frustration indeed make the low energy excitation gapless just as in the Hermitian case. The ground state in this region is not the vacuum state annihilated by all the quasi-particle operators, but the state with one mode occupied. This is in sharp contrast to most non-Hermitian models studied in the literature.

This paper is organized as follows. In \cref{sec:model}, we present the model Hamiltonian and its exact solution. The phase diagram and band structures are illustrated in \cref{sec:band}. Then we discuss the topological properties of the model in \cref{sec:topo}. Finally, we give a conclusion and discussion in \cref{sec:conclu}.

\section{The non-Hermitian XY model}
\label{sec:model}

The Hamiltonian of the anisotropic one-dimensional periodic spin-$1/2$ $XY$ chain with non-Hermitian
symmetric non-collinear interaction in a external magnetic field is given by
\be
&&H=\sum_{j=1}^{L}\left(\frac{1+ \gamma }{2} \sigma_{j}^{x} \sigma_{j+1}^{x}
+ \frac{1- \gamma }{2} \sigma_{j}^{y} \sigma_{j+1}^{y}-h\sigma_{j}^{z}\right) \nonumber\\
&&\quad+\frac{i \delta}{2} \sum_{j=1}^{L} (\sigma_{j}^{x} \sigma_{j+1}^{y} + \sigma_{j}^{y} \sigma_{j+1}^{x})
\label{ham:xynhdm}
\ee
where $\sigma^a$ with $a=x,y,z$ are Pauli matrices. Here $\gamma$ parameterizes the anisotropic couplings, $h$ gives the external field along $z$ direction. The non-Hermiticity arises from the imaginary non-collinear interaction term. The Hermitian version of XY model ($\delta=0$) was solved long time ago by Lieb et al \cite{lieb1961} with the help of Jordan-Wigner transformation \cite{Jordan1928}. The Jordan-Wigner transformation maps the Pauli operators into Majorana fermions operators then the spin model become a quadratic fermion Hamiltonian which is the imbalanced pairing Kitaev chain studied in \cite{Zhang2018}. However, the boundary term contains a extra sign which depends on the fermion number of the system. We will illustrate this point explicitly as follows.

Taking the Jordan-Wigner transformation
\begin{subequations}\begin{align}
\sigma_{j}^{x}&= \left(c_{j}^{\dagger}+c_{j}\right)\exp\Big[i\pi\sum_{l=1}^{j-1}\dc_{l}c_l\Big]\\
\sigma_{j}^{y}&=-i\left(c_{j}^{\dagger}-c_{j}\right)\exp\Big[i\pi\sum_{l=1}^{j-1}\dc_{l}c_l\Big]\\
\sigma_{j}^{z}&= 2 c_{j}^{\dagger} c_{j} - 1
\end{align}\end{subequations}
and after some algebra
%\bes
%\begin{align}
%\sigma_{j}^{x} \sigma_{j+1}^{x} & = c_{j}^{\dagger} c_{j+1} + c_{j+1}^{\dagger} c_{j}
%+ c_{j}^{\dagger} c_{j+1}^{\dagger} + c_{j+1} c_{j}  \\
%\sigma_{j}^{y} \sigma_{j+1}^{y} & = c_{j}^{\dagger} c_{j+1} + c_{j+1}^{\dagger} c_{j}
%- c_{j}^{\dagger} c_{j+1}^{\dagger} - c_{j+1} c_{j}  \\
%i \sigma_{j}^{x} \sigma_{j+1}^{y} & = - c_{j}^{\dagger} c_{j+1} + c_{j+1}^{\dagger} c_{j}
%+ c_{j}^{\dagger} c_{j+1}^{\dagger} - c_{j+1} c_{j}  \\
%i \sigma_{j}^{y} \sigma_{j+1}^{x} & = c_{j}^{\dagger} c_{j+1} - c_{j+1}^{\dagger} c_{j}
%+ c_{j}^{\dagger} c_{j+1}^{\dagger} - c_{j+1} c_{j}
%\end{align}
%\ees
we obtain the imbalanced pairing Kitaev chain
\begin{equation}\bea
H =& \sum_{j=1}^{L-1} c_{j}^{\dagger} c_{j+1} + c_{j+1}^{\dagger} c_{j} +
\Delta_{\alpha} c_{j}^{\dagger} c_{j+1}^{\dagger} + \Delta_{\beta} c_{j+1} c_{j} \\
&-(-1)^{M} \left( c_{L}^{\dagger} c_{1} + c_{1}^{\dagger} c_{L} +
\Delta_{\alpha} c_{L}^{\dagger} c_{1}^{\dagger} + \Delta_{\beta} c_{1} c_{L} \right) \\ &
- h \sum_{j=1}^{L} (2 n_{j} -1)
\label{ham:ipkc}
\eea\end{equation}
Here $c_j$ is fermion annihilation operator at lattice site $j$. For simplicity we have defined $\gamma+\delta=\Delta_{\alpha}$ and $\gamma-\delta=\Delta_{\beta}$ following the same convention as in \cite{Zhang2018}. The present model is different from the Kitaev chain in that the sign of the boundary term
in the second line of \cref{ham:ipkc} depending on the parity of fermion number $M=\sum_{j}^{L} c_{j}^{\dagger} c_{j}$.

When $M \in odd$, we have the periodic boundary condition (PBC) $c_{j+L}=c_{j}$ and when $M \in even$,
we have the anti-periodic boundary condition (APBC) $c_{j+L}=-c_{j}$. After the Fourier transformation
\begin{equation}
c_{q}=\frac{1}{\sqrt{L}} \sum_{j} c_{j} e^{i q j}
\end{equation}
the Hamiltonian becomes
\be
&&H=\sum_q\Big[2(\cos q-h)\dc_q c_q\nonumber\\
&&\quad+i\sin q(\Delta_{\alpha}\dc_{-q}\dc_q+\Delta_{\beta}c_{-q}c_q)+h\Big]
\ee
The possible values of $q$ will depend on the parity of the number of lattice sites and also the parity of the total fermion number. When $L \in Odd$ we have
\bes
\begin{align}
q^{(O, o)}=&\left\{-\frac{L-1}{L} \pi, \ldots,-\frac{2}{L} \pi, 0, \frac{2}{L} \pi, \ldots, \frac{L-1}{L} \pi\right\} \\
q^{(O, e)}=&\left\{-\frac{L-2}{L} \pi, \ldots,-\frac{1}{L} \pi, \frac{1}{L} \pi, \ldots, \frac{L-2}{L} \pi, \pi\right\}
\end{align}
\ees
Here the first superscript index labels the parity of lattice site and the second index labels the parity of fermion number. We will call the case with odd (even) number of fermions as the odd (even) channel.

Similarly, when $L \in Even$ we have
\bes
\begin{align}
q^{(E, o)}=&\left\{-\frac{L-2}{L} \pi, \ldots,-\frac{1}{L} \pi, 0, \frac{1}{L} \pi, \ldots, \frac{L-2}{L} \pi, \pi\right\}\\
q^{(E, e)}=&\left\{-\frac{L-1}{L} \pi, \ldots,-\frac{2}{L} \pi, \frac{2}{L} \pi, \ldots, \frac{L-1}{L} \pi\right\}
\end{align}
\ees
It has been shown that the existence of $q=0$ and $\pi$ mode would strongly affect the ground state properties and correlation behaviors of the spin system \cite{dong2016,dong2017,dong2018,li2019}.

In this paper we will focus on the case $L \in Odd$ first. We introduce the following complex version of Bogoliubov transformation:
\bes\begin{align}
\eta_{q}=& u_{q} c_{q} - i \sqrt{\frac{\Delta_{\alpha}}{\Delta_{\beta}}} v_{q} c_{-q}^{\dagger} \\
\overline{\eta}_{q}=& u_{q} c_{q}^{\dagger} + i \sqrt{\frac{\Delta_{\beta}}{\Delta_{\alpha}}} v_{q} c_{-q}
\end{align}\ees
where the coefficients of the above transformation are
\begin{equation}\bea
u_{q} =& \sqrt{\frac{1}{2}\left(1+\frac{\cos q - h}{\omega(q)}\right)} \\
v_{q} =& \sqrt{\frac{1}{2}\left(1-\frac{\cos q - h}{\omega(q)}\right)}\operatorname{sgn}(\sin q)
\eea\end{equation}
The dispersion of the quasi-particle is given by
\be
\omega(q)=\sqrt{(\cos q - h)^2+\Delta_{\alpha}\Delta_{\beta}\sin^2 q}
\label{eq:om}
\ee
One can easily check that the anti-communication relations are satisfied
\begin{equation}
\left\{ \eta_{q} , \overline{\eta}_{q^\prime} \right\} = \delta_{q,q^\prime}\;,\;
\left\{ \eta_{q} , \eta_{q^\prime} \right\} = \left\{ \overline{\eta}_{q} , \overline{\eta}_{q^\prime} \right\}=0
\end{equation}
Finally the diagonalized Hamiltonian for both odd and even channels are given by
\bes
\begin{align}
H^{(O,o)} & =2 \sideset{}{^\prime}\sum_{q \in q^{(O,o)}} \omega(q) \overline{\eta}_{q} \eta_{q}+\Lambda_{o}\nonumber\\
&+|h-1|+(h-1)-2(h-1) c_{0}^{\dagger} c_{0} \\
H^{(O,e)} & =2 \sideset{}{^\prime}\sum_{q \in q^{(O,e)}} \omega(q) \overline{\eta}_{q} \eta_{q}+\Lambda_{e}\nonumber\\
&+2(h+1)-2(h+1) c_{\pi}^{\dagger} c_{\pi}
\end{align}
\ees
where
\begin{equation}
\Lambda_{e}=-\sum_{q^{(O,e)}} \omega(q)\;,\; \Lambda_{o}=-\sum_{q^{(O,o)}} \omega(q)
\end{equation}
Note that $\overline{\eta}_{q}$ is not a Hermitian conjugate of $\eta_q$, therefore the above Hamiltonian is still non-Hermitian as the original spin model. The full spectrum of the original spin model contains two parts. One part come from the spectra of $H^{(O,o)}$ with odd number of quasi-particles. The other part comes from the spectra of $H^{(O,e)}$ with even number of quasi-particles.

For comparison, we also show the diagonalized Hamiltonian for the case of $L\in Even$ as follows.
\bes
\begin{align}
H^{(E,o)} = & 2 \sideset{}{^\prime}\sum_{q \in q^{(O,o)}} \omega(q) \overline{\eta}_{q} \eta_{q}
+\Lambda_{o}+|h-1|+3+h \nonumber \\ & +2(h-1) c_{0}^{\dagger} c_{0} -2(h+1) c_{\pi}^{\dagger} c_{\pi}\\
H^{(E,e)} = & 2 \sideset{}{^\prime}\sum_{q \in q^{(O,e)}} \omega(q) \overline{\eta}_{q} \eta_{q}
+\Lambda_{e}
\end{align}
\ees
We will see in next section that the low-energy excitation of the spin model is gapless when $L$ is odd and is gapped when $L$ is even.

\section{Phase diagram and band structure}
\label{sec:band}

\begin{figure}
\centering
\includegraphics[width=\columnwidth]{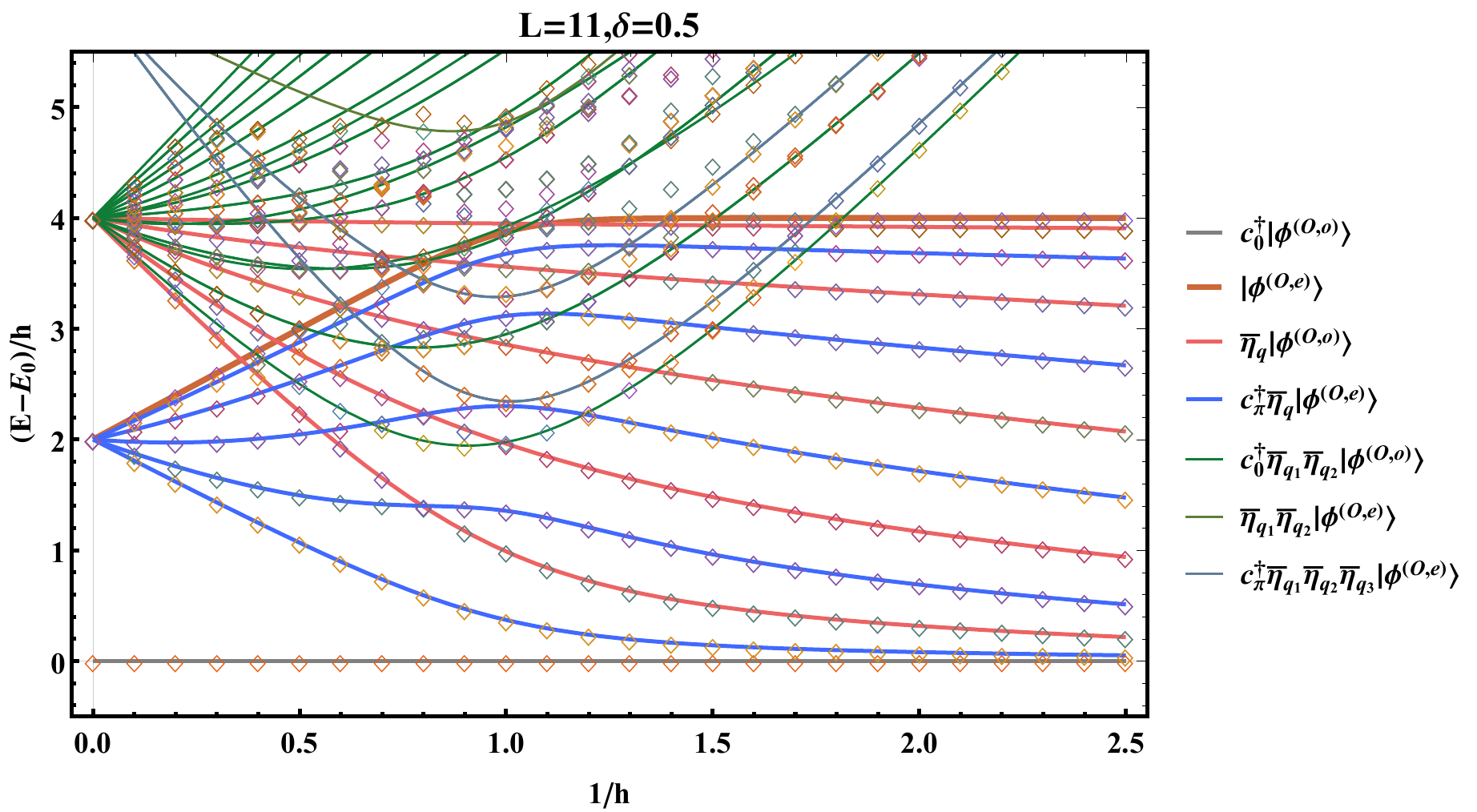}%DM1.eps
\caption{The energy eigenvalues of XY model with non-Hermitian interaction as a function of $1/h$. Here $L=11$ and $\gamma=1$. The dotted diamonds are part of numerical result of exact diagonalization and the solid lines are analytical results. The wave vector $q$ is taken from $q^{(O, o)}$ or $q^{(O, e)}$.}
\label{fig:tek1}
\end{figure}

\begin{figure}
\centering
\includegraphics[width=\columnwidth]{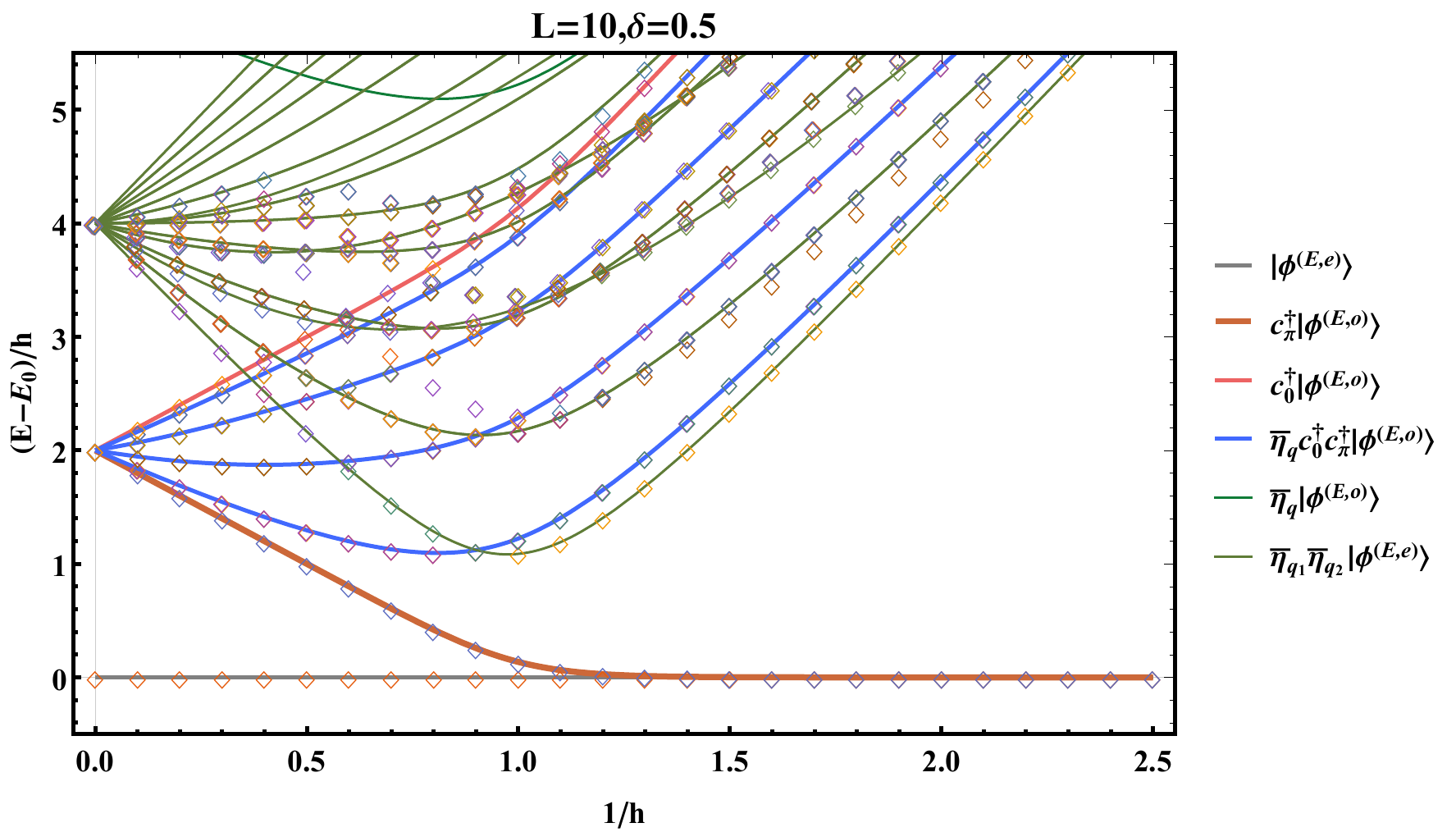}%DM2.eps
\caption{The energy eigenvalues of XY model with non-Hermitian interaction as a function of $1/h$. Here $L=10$ and $\gamma=1$. The dotted diamonds are part of numerical result of exact diagonalization and the solid lines are analytical results. The wave vector $q$ is taken from $q^{(E, o)}$ or $q^{(E, e)}$.}
\label{fig:gap1}
\end{figure}

In this section, we consider the possible phases of our non-Hermitian XY model in details. From Eq.(\ref{eq:om}), it is clear that the eigenvalues are real if the following condition is satisfied.
\begin{equation}
f(q)\equiv(\cos{q}-h)^{2}+\Delta_{\alpha}\Delta_{\beta} \sin^{2}{q}>0 \;,\; -\pi<q<\pi
\label{eq:recond}
\end{equation}
If $|h|<1$, then the first term of $f(q)$ can vanish at certain $q$. In order to achieve $f(q)>0$, we must have $\Delta_{\alpha}\Delta_{\beta}>0$. On the other hand, if $|h|>1$, the first term of $f(q)$ is always positive. In this case, $\Delta_{\alpha}\Delta_{\beta}$ can be negative, and the lowest possible value of $f(q)$ is $\frac{\Delta_{\alpha}\Delta_{\beta}(\Delta_{\alpha}\Delta_{\beta}+h^2-1)}{\Delta_{\alpha}\Delta_{\beta}-1}$. Therefore the condition of $f(q)>0$ is satisfied if we require $\Delta_{\alpha}\Delta_{\beta}+h^2>1$. Other than the above discussed two conditions, the energy eigenvalues are complex.

The region of the real energy eigenvalues in the parameter space is the same as the Kitaev chain with imbalanced pairing in \cite{Zhang2018}. But there is an important difference in the case of $|h|<1$ as we now turn to. First, we note that in this case, the energy spectrum is qualitatively similar for different $\gamma$, therefore we take $\gamma=1$ as an example to illustrate the difference between our model and imbalanced pairing Kitaev chain. Although the vacuum state of the odd channel $\ket{\phi^{(O,o)}}$ is the lowest energy state, but it does not correspond to a valid state in the spin model. The true ground state for the spin model is $\dc_{0}\ket{\phi^{(O,o)}}$, which is a one mode occupied state. Therefore the low energy excitation $\widetilde{\eta}_{q}\ket{\phi^{(O,o)}}$ will approach to the ground state as the system size $L$ become very large, which means the excitation is gapless. Previous works showed that this gapless excitation is make by kink like spin configurations, thus we refer to this region as the kink phase. In \cref{fig:tek1}, we plot the energy eigenvalues of XY model with DMI as a function of $1/h$ for $L=11$. One can see the gapless dispersion appears when $h<1$. This kink phase is actually topological, since the winding number in this region is non-trivial. We will return to this point in next section.

For comparison, we also plot the eigenvalues of XY model with DMI as a function of $1/h$ for $L=10$ in \cref{fig:gap1}. Because of the absence of ring frustration the topological kink phase disappear, the dispersion is gapped except one point at $h=1$, which is same as Kitaev chain \cite{Zhang2018}. In this case the ground states are two-fold degenerated. one round state is $\ket{\phi^{(E,e)}}$, the vacuum state of even channel and the other is  $c_{\pi}^{\dagger} \left\vert \phi^{(E,o)} \right\rangle$ a one mode occupied state from the odd channel.

When $\Delta_{\alpha}\Delta_{\beta}+h^2>1,|h|>1$, the external field is stronger than the spin coupling, the system is in the paramagnetic phase. In this case, excitation is always gapped as can be seen in \cref{fig:tek1} and \ref{fig:gap1}.

In both the kink and the paramagnetic phase, the eigenvalues are real because the model respect the time reversal symmetry ($\mathcal{T}$). In this region, we can construct a Hermitian counterpart \cite{mostafazadeh2005,mostafazadeh2006metric,mostafazadeh2006delta} which reproduce the eigenvalues of the non-Hermitian model in the kink phase. The counterpart can be obtained by replacing the imbalanced paring gaps by their geometrical average.
\be
&&H=\sum_q\Big[2(\cos q-h)\dc_q c_q\nonumber\\
&&\quad+i\sin q(\sqrt{\Delta_{\alpha}\Delta_{\beta}}\dc_{-q}\dc_q
+\sqrt{\Delta_{\alpha}\Delta_{\beta}}c_{-q}c_q)+h\Big]
\ee
In all other regions, the eigenvalues are complex, which indicating that the time reversal symmetry is broken.

\begin{figure}
\centering
\includegraphics[width=0.8\columnwidth]{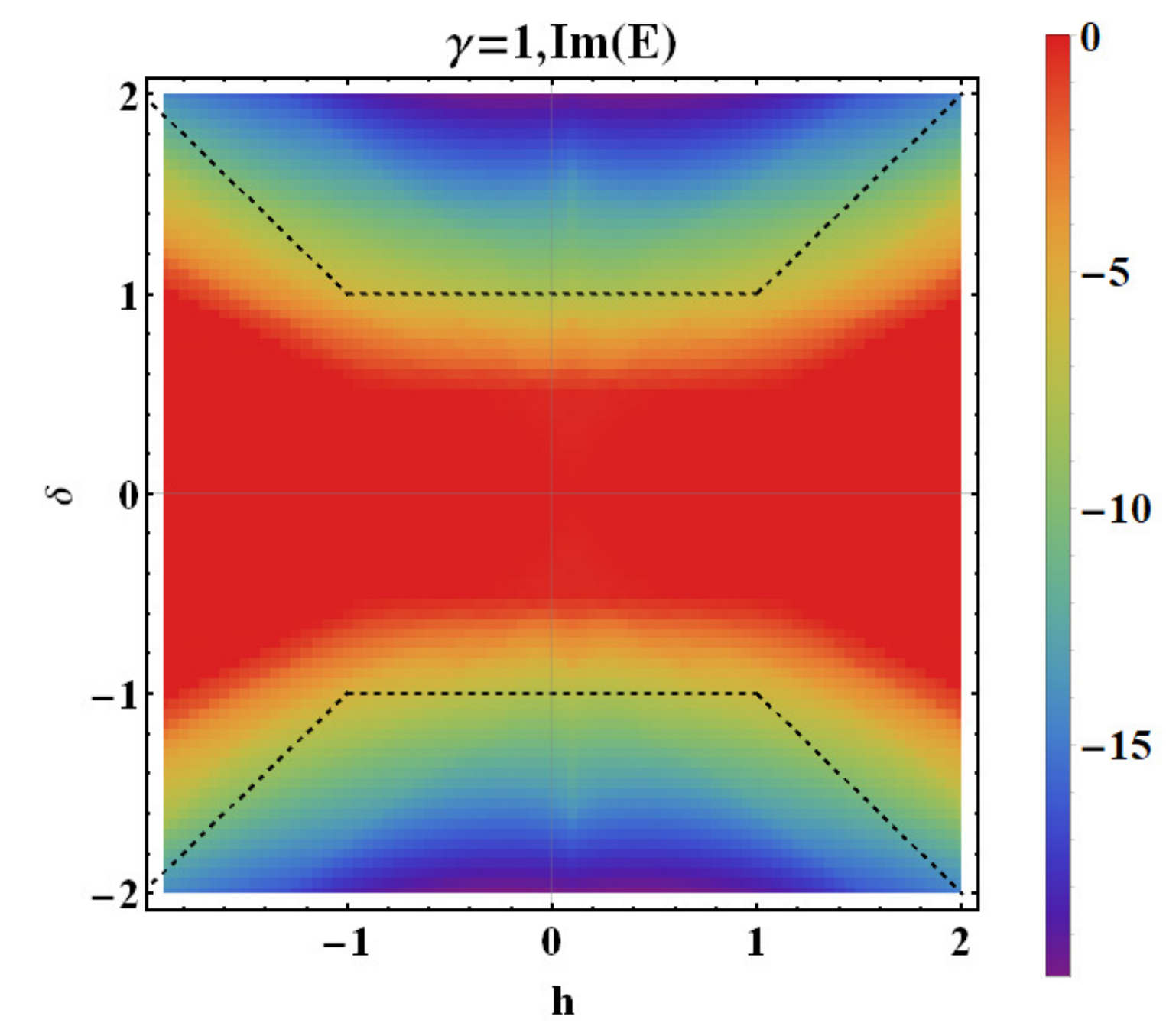}%phase1.eps
\caption{Phase diagram numerically obtained from exact diagonalization of a spin chain of $L=11$.
We take the imaginary part of ground state energy as a criterion. 
The dashed lines are the theoretical phase boundary.
The color bar shows the scale of imaginary part of energy.}
\label{fig:ph1}
\end{figure}

The results of above discussion can be summarized into the \cref{tab}. The third column of the table displays the winding number of each phases, which will be discussed in the next section.
\begin{table}[h!]
\centering
\caption{Phase and topological invariants for $L \in Odd$}\label{tab}
\begin{tabular}{ccc}
\hline
Region                                       & Phase                           & Winding number  \\
\hline
$\Delta_{\alpha},\Delta_{\beta}>0,|h|<1$     & kink         &         1       \\
\hline
$\Delta_{\alpha},\Delta_{\beta}<0,|h|<1$     & kink          &        -1       \\
\hline
$\Delta_{\alpha}\Delta_{\beta}>0,|h|=1$      & critical  &                 \\
\hline
$\Delta_{\alpha}\Delta_{\beta}+h^2>1,|h|>1$  & paramagnetic                 &         0       \\
\hline
Others                                       & $\mathcal{T}$ breaking &                 \\
\hline
\end{tabular}
\end{table}

To verify our result, we use exact diagonalization to obtain the ground state energy.
The phase diagram is present in \cref{fig:ph1}, and we see that the red region in the
parameter plane has real energy eigenvalues, thus $\mathcal{T}$ symmetric. However, the outside
part has complex eigenvalues, and the $\mathcal{T}$ symmetry is breaking. The phase boundary comes
from the result in \cref{tab}.

\section{Topological invariant}
\label{sec:topo}

\begin{figure}
\centering
\includegraphics[width=0.8\columnwidth]{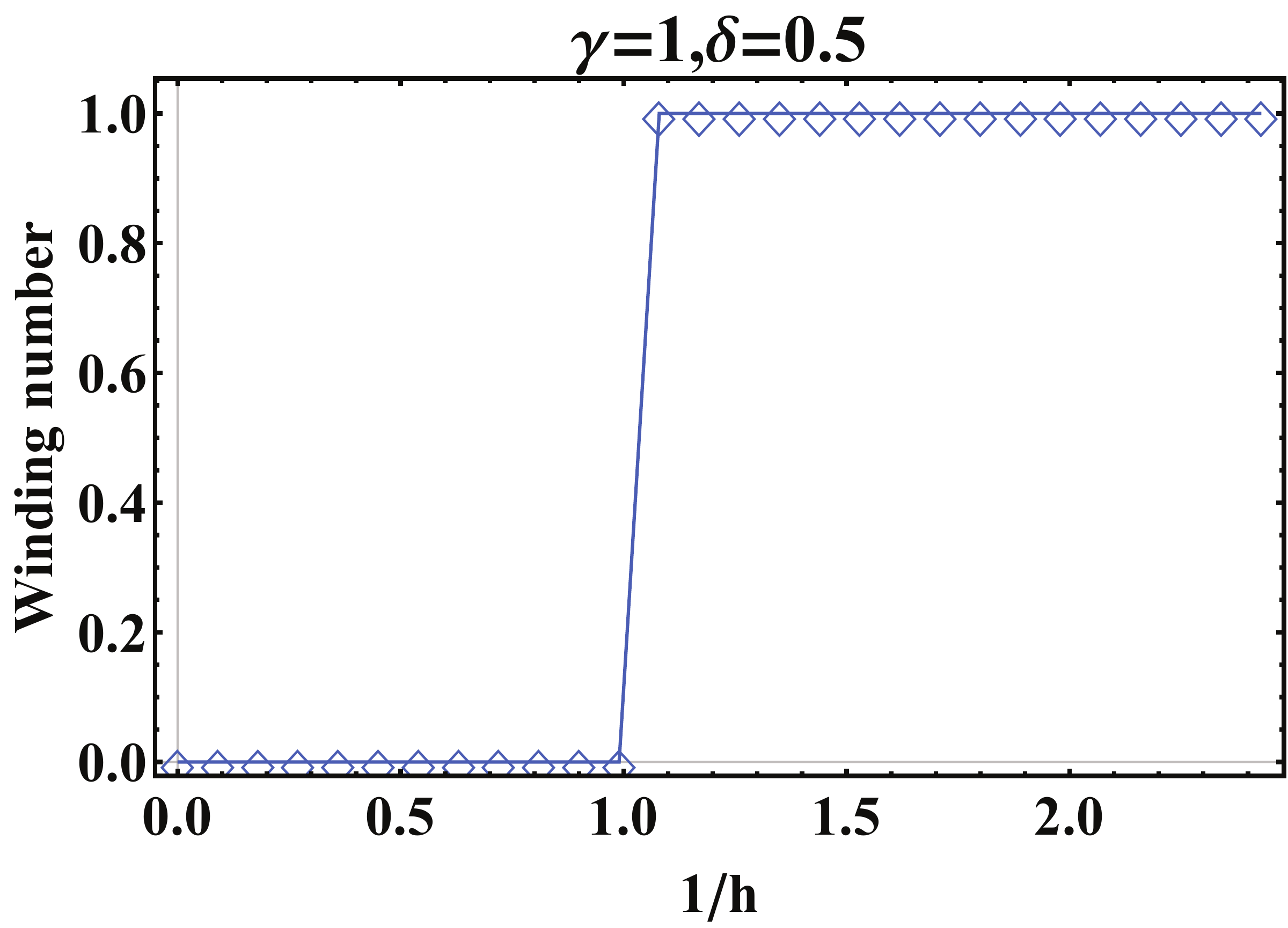}%TN1.eps
\caption{The winding number as a function of $1/h$. We assume $\gamma=1$ and $\delta=0.5$}
\label{fig:tn}
\end{figure}

\begin{figure}
\centering
\includegraphics[width=\columnwidth]{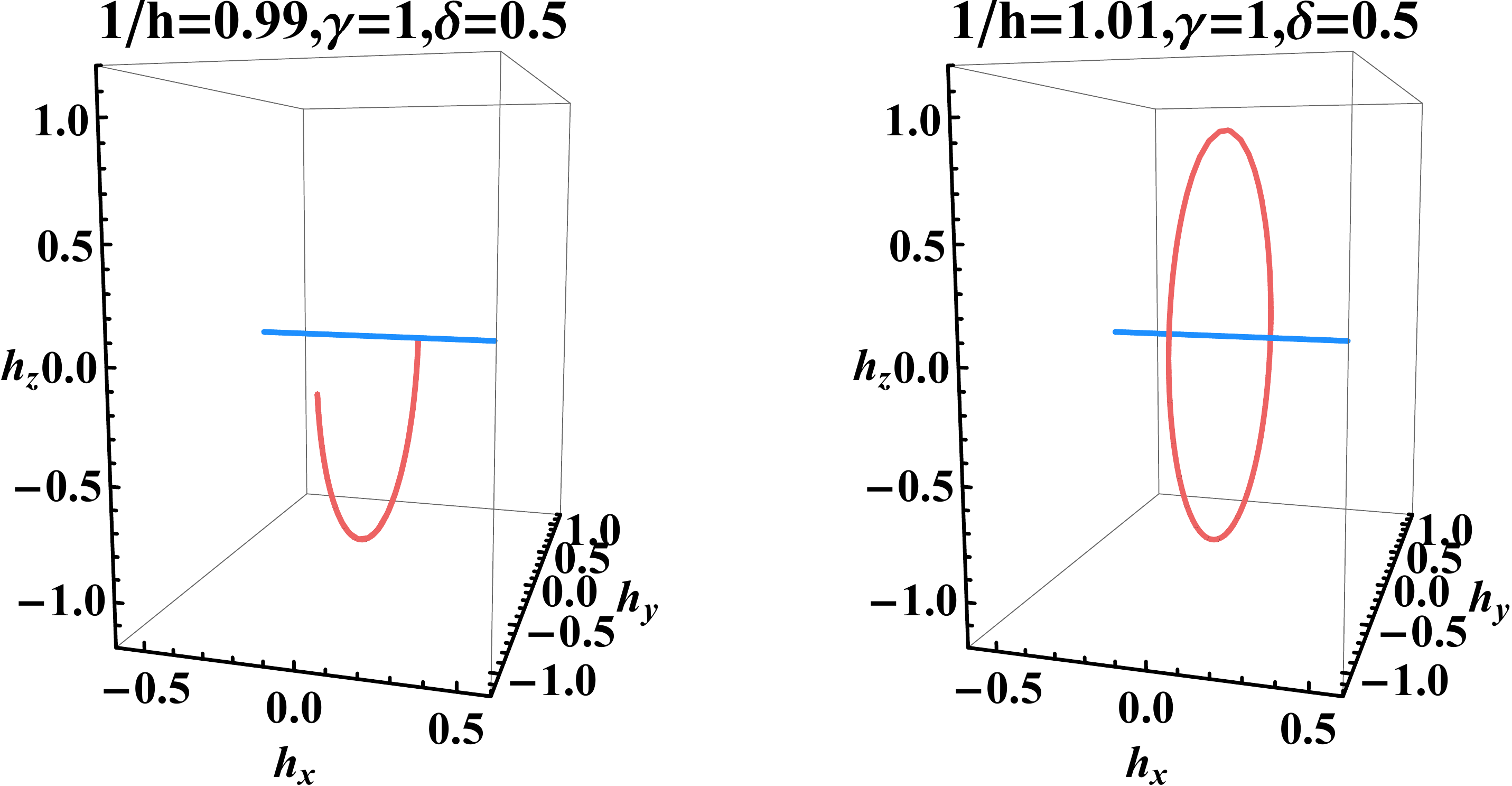}%vech1.eps
\caption{The trajectory of $\mathbf{h}(q)$ in Bloch vector space. The red curve and blue line
denote for the real and imaginary part evolution, respectively. The left panel corresponds
to the trivial case and the right for the topological case.}
\label{fig:h12}
\end{figure}

Now we turn to the topological properties of the model. The Hamiltonian can be
written as $H(q)=\mathbf{h}(q) \cdot \sigma$, where $\mathbf{h}$ is 3-component vector given by
\bes\begin{align}
h_{x}(q) = & - i \delta \sin q \\
h_{y}(q) = &     \gamma \sin q \\
h_{z}(q) = &            \cos q - h
\end{align}\ees
We will refer to the vector $\mathbf{h}$ as the Bloch vector. Normalizing $\mathbf{h}(q)$ as $\widetilde{\mathbf{h}}(q)=\mathbf{h}/|\mathbf{h}|$, the winding number is
defined as
\begin{equation}
w= - \frac{1}{2\pi} \int_{-\pi}^{\pi} \mathrm{d} q \left\vert
\widetilde{\mathbf{h}}(q) \times
\frac{\mathrm{d}}{\mathrm{d} q} \widetilde{\mathbf{h}}(q) \right\vert
\operatorname{sgn}(\Delta_{\alpha})
\end{equation}
The numerical result of winding number as a function of $1/h$ is shown in \cref{fig:tn}. One can see that the winding number jumps at $h=1$, indicating a topological phase transition between the gapless kink phase and the gaped phase.

To visualize the change in the winding number, we plot the trajectory of the vector $\widetilde{\mathbf{h}}(q)$ in the Bloch vector space as $q$ increases from $0$ to $2\pi$ near the topological transition point. In \cref{fig:h12}, we see that the non-Hermitian interaction introduces the
imaginary magnetic flux in the Bloch vector space. When the Bloch vector traces out a closed loop around the flux, the winding number is one. On the other hand, if the trajectory of the Bloch vector does not enclose the flux, it is topologically trivial.

\section{Conclusion}
\label{sec:conclu}

In this paper we have studied the ring frustration in the non-Hermitian situation. Take the paradigmatic $XY$ model as the example, we introduce a non-Hermitian
symmetric non-collinear interaction. The exact solution in periodic boundary condition is obtained with the help of Jordan-Wigner transformation. The analytical solution of energy band matches with that from exact diagonalization perfectly. The gapless kink phase still exists for chains with odd number of sites, as in the Hermitian cases, and we find that the ground state is a single-mode occupied state rather than the vacuum state annihilated by all the quasi-particle operators.
The phase diagram and boundary lines are also presented, with the $\mathcal{T}$ symmetry breaking as the criterion. Finally, we discuss the topological invariant associated with the kink phase. We found a geometrical interpretation of the non-Hermitian interaction in the Bloch vector space, in which the winding number predicts the topological phase transition point.

\acknowledgments

This work is supported by NSFC under Grant No. 11874272.

%----------------------------------------%

\bibliography{ref}

\end{document}